\documentclass[twocolumn,floats,aps,epsfig,nofootinbib,amssymb,preprintnumbers]{revtex4-1}
\usepackage{latexsym,graphicx,epsfig,psfrag,float,amsmath,longtable,amssymb}
\usepackage{cancel,textcomp,bm,times,graphics,hyperref}
\pdfoutput=1
\allowdisplaybreaks
\newcommand{\lsim}{\stackrel{<}{_\sim}}
\newcommand{\gsim}{\stackrel{>}{_\sim}}
\begin{document}
%opening
%\parskip=3pt plus 1pt

%\begin{titlepage}
%\vskip 1cm
%\begin{flushright}
%{
%FTUV-12-XXXX \\
%IFIC-12-
%}
%\end{flushright}
%\vskip 2.5cm

%\setcounter{footnote}{0}
%\renewcommand{\thefootnote}{\fnsymbol{footnote}}

\title{\Large{Zeros of the $W_L Z_L \rightarrow W_L Z_L$ amplitude: where vector resonances stand}}
\author{Alberto Filipuzzi, Jorge Portol\'es and Pedro Ruiz-Femen\'{\i}a}
\affiliation{Departament de F\'{\i}sica Te\`orica, IFIC, CSIC --- Universitat de Val\`encia, Apt. Correus 22085, E-46071 Val\`encia, Spain}
\preprint{FTUV-12-2105}
\preprint{IFIC/12-35}
\date{\today}

%\begin{center}
%{\LARGE \bf Zeros of the $W_L W_L \rightarrow W_L W_L$ amplitude:}
%\\[14pt]{\LARGE \bf where spin-1 resonances stand}
%\vspace{2.5cm} \\
%{\sc  Alberto Filipuzzi, Jorge Portol\'es} and
%{\sc Pedro Ruiz-Femen\'{\i}a}
%\vspace{1.2cm} \\
%$^{b}$ IFIC, CSIC - Universitat de Val\`encia, Apt. Correus 22085, E-46071 Val\`encia, Spain
%\end{center}

%\setcounter{footnote}{0}
%\renewcommand{\thefootnote}{\arabic{footnote}}
%\vspace*{1cm}

\begin{abstract}
A Higgsless electroweak theory may be populated by spin-1 resonances around $E \sim 1 \, \mbox{TeV}$ as a consequence of a new strong interacting
sector, frequently proposed as a tool to smear the high-energy behaviour of scattering amplitudes, for instance, elastic gauge boson scattering.
Information on those resonances, if they exist, must be contained in the low-energy couplings of the electroweak chiral effective theory. Using the
facts that: i) the scattering of longitudinal gauge bosons, $W_L, Z_L$, can be well described in the high-energy region ($E \gg M_W$) by the
scattering of the corresponding Goldstone bosons (equivalence theorem) and ii) the zeros of the scattering amplitude carry the information on the
heavier spectrum that has been integrated out; we employ the ${\cal O}(p^4)$ electroweak chiral Lagrangian to identify
the parameter space region
of the low-energy couplings where vector resonances may arise. An estimate of their masses is also provided by our method.
\end{abstract}

%\vfill
%PACS~:  \\
%Keywords~: Chiral symmetry, Effective Lagrangian.

%\end{titlepage}
\maketitle

\section{Introduction}
\label{sec:1}
LHC will conclude soon the search for the Standard Model (SM) Higgs boson below the TeV energy region. As up to now,
with only a tiny open window of $\sim 20 \, \mbox{GeV}$, there are good chances that there will be no Higgs.
If this is the case the search for the dynamics of the spontaneous symmetry breaking of the electroweak gauge symmetry will become a crucial goal of the
high-energy physics research.
\par
A Higgsless world would be most probably characterized by the presence of a new physics scale associated to a strong interacting sector lying
around $E \sim 1 \,\mbox{TeV}$ and related with the spontaneous breaking of the electroweak symmetry \cite{Veltman:1976rt,Lee:1977yc}.
A reasonable assumption, based on our knowledge of low-energy hadron physics, is that such a non-perturbative dynamics would lead to resonances
that can be at the reach of future runs at the LHC or at an envisaged Linear Collider. It has also
been argued that without a light Higgs the violation of perturbative partial-wave unitarity in the elastic scattering of the longitudinal
components of those W or Z gauge bosons could be prevented by those spin-1 resonances \cite{Veltman:1991cm}
though this is by no means compulsory \cite{Lee:1977yc}. In this article we plan to investigate the possible existence of those vector resonances
contributing to the $W_L Z_L \rightarrow W_L Z_L$ scattering process.
\par
 The symmetry breaking sector of the Standard Model without a Higgs becomes a non-linear sigma model with $SU(2)_L \otimes SU(2)_R / SU(2)_V$ symmetry
where the $SU(2)_L \otimes U(1)_Y$ gauge symmetry is properly embedded. Interestingly enough the Lagrangian that describes it is the one of
two-flavour Chiral Perturbation Theory (ChPT) \cite{Weinberg:1978kz,Gasser:1983yg} with pions substituted by the Goldstone bosons
that provide masses to the gauge bosons. ChPT is an effective field theory of Quantum Chromodynamics (QCD) at very low-energy,
driven by the chiral symmetry of massless QCD, perturbative in momenta and valid for $p^2 \ll (4 \pi F_{\pi})^2$
(being $F_{\pi}$ the decay constant of the pion), and renormalizable order
by order in its perturbative expansion.
\par
As in any effective field theory the low-energy coupling constants (LECs) of ChPT carry the information of the heavier spectra that has been left
out in the procedure of constructing the low-energy theory. Indeed it has been shown that, at ${\cal O}(p^4)$, the LECs are saturated by the
lightest hadron resonances \cite{Ecker:1988te}. In particular the two relevant LECs that appear in the amplitude of the elastic pion-pion scattering are given
by the contribution of the $\rho(770)$. Hence one can wonder if it would be possible to obtain the mass of the $\rho(770)$ from the values of
those LECs. Of course it is not possible to establish the existence and properties of resonances using a
perturbative framework. However one can provide procedures to resummate the perturbative contributions to the amplitude.
We will translate one of these methods from the well studied QCD framework to the electroweak sector.
\par
The procedure that we devise in order to explore the occurrence of spin-1 resonances in the $E \sim 1 \, \mbox{TeV}$ region is
based on the information about the spin-J resonances ($J \ge 1$) provided by the zeros of the scattering amplitude.
This method goes back to the study of the zeros in $\pi \pi \rightarrow \pi \pi$~\cite{Pennington:1973dz}.
In Ref.~\cite{Pennington:1994kc} it was shown in the framework of ChPT that the zeros of the $I=1$ $\pi \pi \rightarrow \pi \pi$ amplitude
($I$ is short for the isospin quantum number) predict the mass of the $\rho(770)$ resonance when the chiral LECs are saturated by the
resonance contributions. This shows that, though the ChPT amplitude is only valid for $p^2 \ll M_{\rho}^2$, the extrapolation provided by 
its zeros is to be trusted up to $E \sim M_{\rho}$.
\par
The method can be applied to the electroweak sector employing
the two following ingredients. First, the fact that the elastic scattering amplitude of the longitudinal components of the gauge bosons is given,
at $E \gg M_W$, by the amplitude of the
elastic scattering of the Goldstone bosons associated to the spontaneous electroweak symmetry breaking. This is known as the {\em equivalence theorem}
and was devised originally to study those processes with a very heavy Higgs \cite{Cornwall:1974km,Lee:1977yc}.
The equivalence theorem thus allow us to trade the dynamics of the longitudinally polarized gauge bosons by the one of the corresponding
Goldstone modes. And the second ingredient, already commented above, is the fact that the interactions among Goldstone bosons in the
Higgsless electroweak theory is described, at least at leading order, by the two-flavour ChPT Lagrangian where now the multiplet of
pions is substituted by the Goldstone fields that provide masses to the gauge bosons. The obvious difference is the relevant
scale that rules the perturbative expansion of the amplitude \cite{Longhitano:1980iz,Longhitano:1980tm,Appelquist:1993ka}. Indeed the perturbative
scale is now driven by $v \sim (\sqrt{2} \, G_F)^{-1/2} \simeq 246 \, \mbox{GeV}$ with $G_F$ the Fermi constant. Accordingly the effective
theory is  valid for $p^2 \ll (4 \pi v)^2 \sim \left( 3 \, \mbox{TeV} \right)^2$.  Taking into account the equivalence theorem,
the perturbative expansion and the Higgsless electroweak chiral effective theory (EChET) just mentioned, our working region is
determined by $M_W \ll E \ll 4 \pi v$.
\par
 We therefore exploit in this work the analogies between ChPT and Higgsless EChET and study the elastic scattering of the longitudinal components of the
electroweak gauge bosons as described by the latter. Then, assuming that the LECs of the effective theory are saturated by the lightest
vector resonances (if these exist) we determine the zeros of the amplitude and we explore the parameter space of the
two only LECs that appear in the scattering amplitude, obtaining important information on the possible resonances.
\par
The contents of this article are the following.
In Section II we revisit the role of the zeros of an amplitude and its relation with the resonances of the theory. We will
focus on the well known case of $\pi \pi \rightarrow \pi \pi$ scattering and the $\rho(770)$.
In Section III we briefly review the Higgsless electroweak effective theory applied
to the $W_L Z_L \rightarrow W_L Z_L$ scattering on which we will particularize our study.
Section IV will be devoted to the analysis of the zeros of that amplitude and
their interpretation as vector resonances. We will also discuss our results and the
possibility left for LHC to disentangle the presence of these
vector resonances. Our conclusions will be given in Section V. An Appendix reminds the reader the general conditions of convergence
of the partial-wave expansion of the elastic $\pi\pi$ scattering amplitude.

\section{The role of the zeros of the scattering amplitude}
\label{sec:2}

We will now develop our method using QCD as the reference framework, taking advantage of the precise experimental data available and the
good knowledge of the LEC values of $\mathcal{O}(p^4)$ ChPT used to describe the low-energy processes like $\pi\pi$ scattering. No conceptual changes will be needed to apply the method to the electroweak case.
\par
The low-energy dynamics of elastic $\pi \pi$ scattering is determined by the existence of the lightest meson resonances contributing to that
amplitude, $\sigma (600)$ and $\rho(770)$. Though the $\sigma (600)$ is mainly related with the chiral logs (it appears at next-to-leading
order in the large number of colours ($N_C$) expansion), the information of the $\rho(770)$, leading in $1/N_C$, is encoded in the low-energy
couplings at ${\cal O}(p^4)$ in the chiral expansion
\cite{Ecker:1988te}. Within a quantum field theory approach, one can determine the resonance contributions to the chiral LECs starting from a
Lagrangian with explicit resonance fields and then integrating them
out. Typically those LECs are given in terms of the resonance couplings to the pions and inverse powers of the resonance masses.
\par
One can wonder if the opposite procedure is viable. That is, if it would be possible to determine the mass
of the resonances from the phenomenological determination of the LECs. As the ChPT amplitudes provide a perturbative expansion in momenta
it is clear that the poles of resonances are not a feature of chiral symmetry. However a link between chiral dynamics and resonance
contributions can be provided employing some ad-hoc resummation techniques like Pad\'e approximants, the inverse amplitude method or the $N/D$
construction \cite{MartinSpearman:1970,MartinMorgan:1976,Dobado:1997jx}. We will propose an alternative procedure based on the zeros of the scattering
amplitude~\cite{Pennington:1971fd,Pennington:1973dz,Pennington:1973xv,Pennington:1973hi,Pennington:1994kc} as given by ChPT at ${\cal O}(p^4)$
(we will also comment on the next ${\cal O}(p^6)$ contribution).
\par
Consider the amplitude $F(s,t)$ for $\pi^-(p_1) \pi^0(p_2) \rightarrow \pi^- \pi^0$ in the s-channel:
\begin{equation} \label{eq:st}
  s=(p_1+p_2)^2, \; \; \; \; \; \; \;  t = \frac{1}{2} (s-4M_{\pi}^2)(\cos \theta-1).
\end{equation}
This amplitude has no $I=0$ component, and from the phenomenology we know that the isovector P-wave is large
whereas the $I=2$ (exotic) S-wave is small.
We can anticipate that these features are essential for our method.
The P-wave is dominated by the $\rho(770)$ resonance and therefore around this energy region we can write the partial-wave expansion
of the amplitude as:
\begin{equation} \label{eq:pwfst}
 F(s,t) = 16\pi f_0^2(s) + \frac{48\pi}{\sigma}  \frac{M_{\rho} \Gamma_{\rho}(s)}{M_{\rho}^2-s-i M_{\rho} \Gamma_{\rho}(s)} \, \cos \theta
+\,\dots,
\end{equation}
where $\sigma = \sqrt{1-4M_{\pi}^2/s}$  and $f_{\ell}^I(s)$ is the partial-wave with isospin $I$ and angular momentum $\ell$, defined through
the partial-wave expansion of the s-channel amplitude with defined isospin:
\begin{equation} \label{eq:pw}
 F^I(s,t) = 32\pi \sum_{\ell = 0}^{\infty} (2 \ell + 1) f_{\ell}^I(s) P_{\ell}(\cos \theta) \, ,
\end{equation}
with $P_{\ell}(z)$ the Legendre polynomial of  degree $\ell$. 
Unitarity imposes severe constraints on the structure of the partial-waves
$f_{\ell}^I(s)$. A description consistent with unitarity is given by
\begin{equation} \label{eq:phase}
 f_{\ell}^I(s) = \frac{1}{\sigma} \, e^{i \delta_{\ell}^I} \sin \delta_{\ell}^I ,
\end{equation}
with $\delta_{\ell}^I$ the phase shift of isospin $I$ and angular momentum $\ell$, that is real for elastic scattering.
\par
The remaining terms not quoted in Eq.~(\ref{eq:pwfst}) amount to numerically suppressed higher partial waves.
Taking into account the small size of the S-wave component,
the angular distribution associated to $F(s,t)$ would have a marked dip at $\cos \theta = 0$, where also $F(s,t)\simeq 0$. This reflects the spin-1
nature of the $\rho(770)$. Due to the properties of the Legendre polynomials these dips in the angular distribution (or zeros of the amplitude)
will appear for $\ell > 0$ and their number in
the physical region,  $\cos \theta \in [-1,1]$, will be given by the angular momentum of the partial-wave. These zeros can be considered
as dynamical features which give the spin to the resonance.
\par
This observation gives us a possible path to analyze the spectrum of $J \geq 1$ resonances integrated out and hidden in the couplings of the effective
field theory. Let us specify  several features of the zeros of the amplitude and give precise definitions that will help to fix our procedure.
\par
Being analytical functions of more than one variable the zeros of the amplitude are not isolated but continuous, defining a one-dimensional manifold for real $s$ and complex $t$.
Using Eq.~(\ref{eq:st}) the $F(s,t)$ amplitude in the s-channel may be expressed as $F(s,z)$ with $z \equiv \cos \theta$.
Then the solution of $F(s,z_0) = 0$ for physical values of the $s$ variable is defined by
$z = z_0(s)$. Though the zeros of the function happen at complex values of the $z$ variable, we define the {\em zero contour} as the
real part of the zeros ($\mbox{Re}\,z_0(s)$).
It is also phenomenologically observed \cite{Odorico:1972eb,Barrelet:1971pw} that this contour continues smoothly from one region to another in the
Mandelstam plane\footnote{There is one known situation where this is not the case and the contours wiggle. This happens when they pass a threshold that opens
strongly in the S-wave. One such example is the $K \overline{K}$ threshold in $I=0$ $\pi \pi$ scattering \cite{Pennington:1972zp}. As we consider here the $\pi^- \pi^0$ channel,
which has no isoscalar component, we expect the zero contour to be quite smooth well beyond the $\rho(770)$ mass.}.
Using Eqs.~(\ref{eq:pwfst},\ref{eq:phase}) we get:
\begin{equation} \label{eq:mikes}
 z_0(s) = - \frac{e^{i \delta^2_0} \sin \delta^2_0}{3 \, M_{\rho} \, \Gamma_{\rho}(s)} \left( M_{\rho}^2 - s - i M_{\rho} \Gamma_{\rho}(s) \right) \, .
\end{equation}
In the narrow resonance approximation (dropping $\Gamma_{\rho}(s)$ in the numerator of Eq.~(\ref{eq:mikes})) we see that $z_0(M_{\rho}^2)=0$.
For a finite $\rho(770)$ width we have:
\begin{equation} \label{eq:mikep}
 \mbox{Re}\,z_0(s) = - \frac{\sin 2 \delta^2_0}{6 \, M_{\rho} \, \Gamma_{\rho}(s)} \left( M_{\rho}^2 - s \right) - \frac{1}{3} \sin^2 \delta_0^2 \, ,
\end{equation}
that satisfies $\left| \mbox{Re}\,z(M_{\rho}^2) \right| \leq \frac{1}{3}$. In fact, and due to the exotic character of the
S-wave $I=2$ background $\left(\mbox{Im} f_0^2(s) \ll \mbox{Re} f_0^2(s)\right)$ and to the absence of the S-wave $I=0$ channel,
we know that $\left| \mbox{Re}\,z(M_{\rho}^2) \right| \ll\frac{1}{3}$.
\par
Hence, for a generic amplitude where the P-wave contribution dominates and is saturated by a vector resonance, the resonance mass $M_R$
should be found as the solution of:
\begin{equation} \label{eq:master}
\mbox{Re}\,z_0(M_R^2) \simeq 0 \, ,
\end{equation}
where $z_0(s)$ is the zero contour obtained from that amplitude.
We will take this condition as our source of information on the resonances given by $F(s,t)$.
It is clear, from Eq.~(\ref{eq:pwfst}), that how deep is the dip of the angular distribution will also depend on the size of the imaginary
part of the zeros. We will comment on this point further in Section \ref{sec:4}.
\par
\begin{figure}[tb]
\begin{center}
\includegraphics[width=8.5cm]{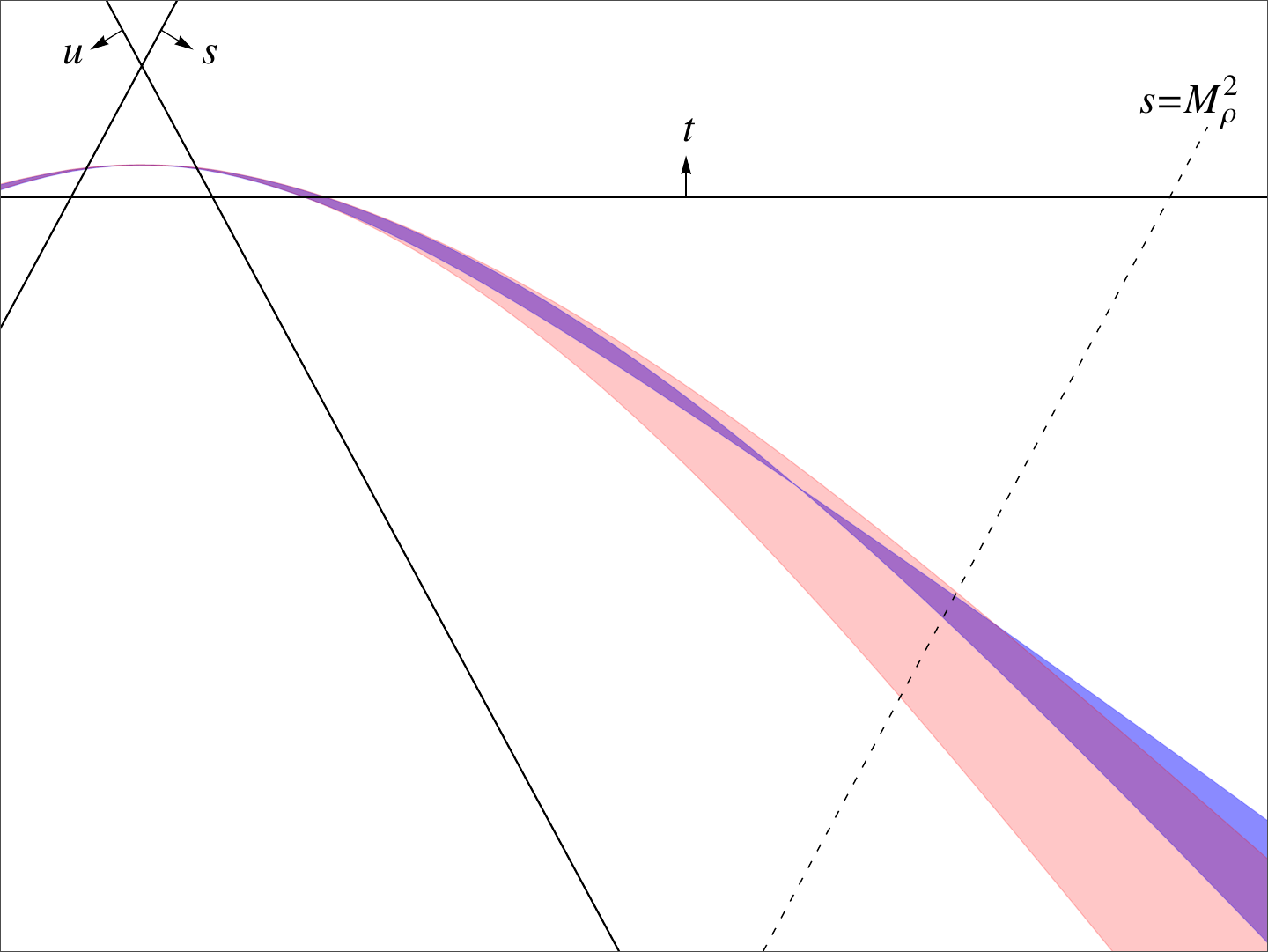}
\caption{\label{fig:1} Zero contours of $\pi^- \pi^0 \rightarrow \pi^- \pi^0$ at ${\cal O}(p^4)$
(light colour) and ${\cal O}(p^6)$ (dark colour) in the Mandelstam plane. The bands correspond to the range of LECs for
 $\mu= 0.6 \, \mbox{GeV}$ and $\mu = 0.9 \, \mbox{GeV}$ as given in Table~\ref{tab:1}.
The zero contours connect with the Weinberg's projection of the Adler zero inside the Mandelstam triangle \cite{Weinberg:1966kf}.}
\end{center}
\end{figure}

Let us repeat (and update) now the procedure in Ref.~\cite{Pennington:1994kc}, showing how the method developed works in a real case. The ${\cal O}(p^4)$ amplitude of $\pi^- \pi^0 \rightarrow \pi^- \pi^0$ , with the variables defined in Eq.~(\ref{eq:st}), is given by \cite{Gasser:1983yg}:
\begin{widetext}
\begin{eqnarray} \label{eq:op4pp}
A(t,s,u) &=&\frac{t-M^2}{F^2} +\frac{1}{6\,F^4} \bigg[ 3(t^2-M^2) \, \bar{J}(t)
 + [ s(s-u)-2\,M^2\,s+4\,M^2\,u-2\,M^4] \, \bar{J}(s) \nonumber  \\[3mm]
&& \qquad \qquad \qquad \quad  + [u(u-s)-2\,M^2\,u+4\,M^2\,s-2\,M^4] \, \bar{J}(u) \bigg]  \\[3mm]
&& + \frac{1}{96 \pi^2\,F^4} \left[ 2 \left( \overline{\ell}_1 - \frac{4}{3} \right)(t-2\,M^2\,)^2
+  \left( \overline{\ell}_2 - \frac{5}{6} \right) (t^2+(s-u)^2) -12\,M^2\,t+15\,M^4 \right] , \nonumber
\end{eqnarray}
\end{widetext}
where $F$ is the decay constant of the pion in the chiral limit, $F \simeq F_{\pi} \simeq 92.4 \, \mbox{MeV}$ and
$M\simeq M_\pi \simeq 138 \, \mbox{MeV}$. The one-loop $\bar{J}(x)$ function was defined in Ref.~\cite{Gasser:1983yg}.
$\overline{\ell}_1$ and $\overline{\ell}_2$ are a priori unknown low-energy couplings related with those of the ${\cal O}(p^4)$ ChPT
Lagrangian with two flavours, $\ell_i^r(\mu)\,(i=1,2)$, through:
\begin{eqnarray} \label{eq:lbar}
 \ell_1^r(\mu) &=& \frac{1}{96 \pi^2} \left( \overline{\ell}_1 + \ln \frac{M_{\pi}^2}{\mu^2} \right) \, , \nonumber \\
\ell_2^r(\mu) &=& \frac{1}{48 \pi^2} \left( \overline{\ell}_2 + \ln \frac{M_{\pi}^2}{\mu^2} \right) .
\end{eqnarray}
Here $\mu$ is the renormalization scale. With these definitions the $\overline{\ell}_i$ couplings are, but for a factor,
equal to $\ell_i^r(M_{\pi}^2)$ and thus scale independent. It is well known \cite{Ecker:1988te} that the ${\cal O}(p^4)$ chiral LECs
are saturated by the contribution
of the lightest resonances that have been integrated out. In fact, $\ell_1^r(\mu)$ and $\ell_2^r(\mu)$ are saturated by the lightest multiplet of vector
resonances. Upon resonance integration at tree level one gets the
three-flavour LECs $L_i^r(\mu)$. Using the ${\cal O}(p^4)$ matching with the two-flavour $\ell_i^r(\mu)$ \cite{Gasser:1984gg},
the resonance contributions to the latter read $\ell_1^r(\mu) = -G_V^2/M_V^2-\nu_K/24$ and $\ell_2^r(\mu)=G_V^2/M_V^2-\nu_K/12$. Here $M_V\simeq M_\rho$ is the
mass of the lightest nonet of vector resonances, and $G_V$ is a coupling of the Resonance Chiral
Theory phenomenological Lagrangian \cite{Ecker:1988te}.
\begin{table}[tb]
\begin{tabular}{|c|c|c|c|}
\hline
$\mu (\mbox{GeV})$ & $0.6$ & $0.77$ & $0.9$ \\
\hline
$\overline{\ell}_1$ & $-0.33$ & 0.25 & $0.58$ \\
$\overline{\ell}_2$ & $4.46$ & 5.03  & $5.37$ \\
$10^5 r_5^V$ & $5.55$ & $4.96$ & $4.78$ \\
$10^5 r_6^V$ & $0.67$ & $0.86$ & $0.94$ \\
\hline
\end{tabular}
\caption{\label{tab:1} Renormalization scale dependence of the LECs determinations relevant for the evaluation of the zero contours at
${\cal O}(p^4)$ and ${\cal O}(p^6)$ in ChPT. }
\end{table}
Its value is estimated as $G_V \in [40,50] \, \mbox{MeV}$ \cite{Pich:2010sm,Dumm:2009va} and we take $G_V \simeq 45 \, \mbox{MeV}$ for the numerical
evaluation.
In addition $\nu_K = \left(\ln \left( M_K^2/\mu^2\right) +1\right) /(32 \pi^2)$.
The constants $\overline{\ell}_i$ become $\mu-$dependent if we substitute the $\ell_i^r(\mu)$ in Eq.~(\ref{eq:lbar}) by the tree-level estimates above.
It is generally assumed that vector resonance saturation of the low-energy constants implies that the resonance contributions
determine the LECs quite well for a scale $\mu$ of the order of the mass of the resonance. Within the interval $\mu \in [0.6,0.9]$~GeV,
the couplings $\overline{\ell}_1$ and  $\overline{\ell}_2$ take the values shown in Table~\ref{tab:1}; for the central value, $\mu=M_{\rho}$,
one gets $\overline{\ell}_1 = 0.25$, $\overline{\ell}_2 = 5.03$. With these estimates, we can readily evaluate the zero contour from the
${\cal O}(p^4)$ ChPT, and obtain the $\rho(770)$ mass through Eq.~(\ref{eq:master}). In the Mandelstam plane the outcome for the zero contours
takes the form of the lighter band in Figure~\ref{fig:1}, whose
limits are given by the extreme values of $\overline{\ell}_1$ and $\overline{\ell}_2$ in Table~\ref{tab:1}. These zero contours intersect
the Re~$z$-axis for $M_R\in [0.69,0.91]$~GeV. The estimate is particularly good for the central value $\mu=M_\rho$, which yields
$M_R= 0.75$~GeV. Let us note that the $\mu$-dependence of the zero contour prediction for the $\rho(770)$ mass
does not reflect any uncertainty intrinsic to the method, but is just a consequence of the incomplete knowledge of the resonance estimates
of the LECs.
\par
We would like next to show how the zero contour in Figure~\ref{fig:1} changes when the next chiral order is considered.
The $\pi^- \pi^0 \rightarrow \pi^- \pi^0$ amplitude up to ${\cal O}(p^6)$ has been computed in
Ref.~\cite{Bijnens:1997vq}. At this order the result
depends on 6 low-energy couplings $r_i^r(\mu)$, $i=1,\dots 6$. Assuming that the values of the latter are saturated by vector resonance contributions,
one can get estimates for $r_i^r(\mu)\simeq r_i^V$~\cite{Bijnens:1997vq} in the same way as for the $\ell_i^r(\mu)$. We take the values
of $r_i^V, i=1,2,3,4$, as given in \cite{Bijnens:1997vq}. However we have an updated evaluation for $i=5,6$, obtained
as follows: we use the definition of $r_5^V$ and $r_6^V$ in terms of
the ${\cal O}(p^6)$ two-flavour LECs $c_i^r(\mu)$ \cite{Bijnens:1999hw},
then the relation of the two-flavour LECs with the three-flavour ones $C_i^r(\mu)$~\cite{Gasser:2009hr}, and
finally we determine the latter by integrating the resonance fields at tree-level~\cite{Cirigliano:2006hb}.
Incidentally only vector resonances contribute to $r_5^V$ and
$r_6^V$. The numerical values of $r_5^V$ and $r_6^V$ that we get, shown in Table~\ref{tab:1},
are typically a factor of 2 to 4 smaller than the ones quoted in Ref.~\cite{Bijnens:1997vq}.
This procedure cannot be applied, at the moment, for $r_i^V$, $i=1,2,3,4$, because the conversion between the corresponding
LECs with two and three flavours has not been made available yet.
The obtained ${\cal O}(p^6)$ zero contours are displayed by the darker band of Figure~\ref{fig:1}, that is defined
by the $\mu$-dependent values of the LECs given in Table~\ref{tab:1}. The $\rho(770)$ mass estimates from these
${\cal O}(p^6)$ zero contours read $M_R\in [0.83,1.01]$~GeV. 
We notice that the resonance estimate of $r_4^V$ involves a cancellation of two different contributions of similar size.
If we, for instance, change the sign of $r_4^V$ the $\rho(770)$ mass estimates at ${\cal O}(p^6)$ are shifted to $M_R\in [0.79,0.90]$~GeV.
The strong dependence of these determination on the values
of the $r_i^r(\mu)$ (specially on $r_4^r(\mu)$ and $r_6^r(\mu)$) calls for an updated evaluation of the latter
before assessing the validity of the zero contour approach at ${\cal O}(p^6)$ for the case of the $\rho(770)$.
\par
The zero contours provide also an alternative unitarization procedure for the ChPT amplitude. Indeed assuming that the amplitude
satisfies the partial-wave expansion (\ref{eq:pw}) and
neglecting $\ell \geq 2$ partial waves one obtains:
\begin{equation} \label{eq:pedrodia}
 \tan \delta_1^1(s) = \frac{-\frac{1}{2} \sin 2 \delta_0^2(s)}{3\, \mbox{Re}\,z_0(s) + \sin^2 \delta_0^2(s)} \, .
\end{equation}
Note that, if the S-wave $I=2$ phase-shift is small, the P-wave phase shift
$\delta_1^1$ will pass through $\pi/2$ when $\mbox{Re}\,z_0(s) \rightarrow 0$, thus
indicating the presence of a resonance. The zero contour provided by the  ${\cal O}(p^4)$ ChPT amplitude could then be
employed to obtain the contribution of the lightest vector resonance (the $\rho(770)$ in the QCD case) to the $\delta_1^1(s)$
phase shift. This is by no means evident. The chiral expansion provides an accurate description at very low energies only,
namely for $E \ll M_{\rho}$. Hence the fact that the zero contour is able to unitarize the theory at $E \sim M_{\rho}$ has
to rely on the properties of those zero contours.
\begin{figure}[t]
\begin{center}
\includegraphics[width=8cm]{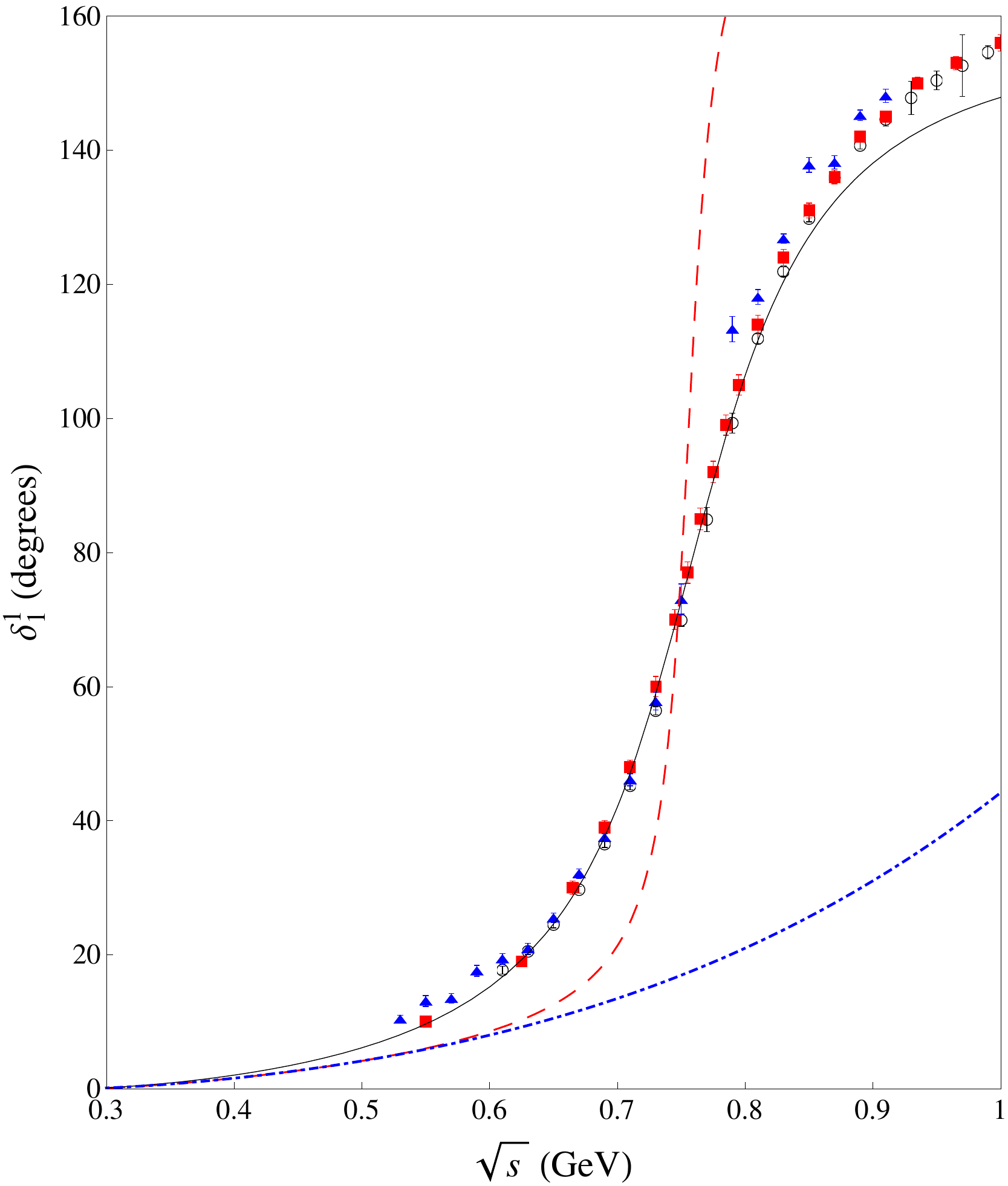}
\caption{\label{fig:2} $\delta_1^1(s)$ phase-shift in elastic $\pi \pi$ scattering. Comparison of experimental
data~\cite{Protopopescu:1973sh,Ochs:1973,Estabrooks:1974vu}  with the
 prediction given by Eq.~(\ref{eq:pedrodia}) for
 the S-wave given by the Schenk parameterization (continuous line) and by ${\cal O}(p^4)$ ChPT (dashed line).
The ${\cal O}(p^4)$ ChPT result for the P-wave phase-shift, namely $\delta_1^1(s)= \sigma \mbox{Re} f_1^1(s)$, is also shown (dot-dashed line).}
\end{center}
\end{figure}
\par
In Figure~\ref{fig:2} we compare the experimental data on the $\delta_1^1(s)$ phase-shift from  elastic
$\pi \pi$ scattering with different theoretical predictions. As expected the prediction given by the ChPT amplitude does not provide
the right description for $E \sim M_{\rho}$, as can be seen looking at the dot dashed line that represents the ChPT result for the P-wave phase-shift, namely
$\delta_1^1(s) = \sigma \, {\mbox Re} f_1^1(s)$~\cite{Gasser:1990ku}. On the other hand the $\delta_1^1(s)$ phase-shift as given
by (\ref{eq:pedrodia}) is in much better agreement with experimental data. We plotted the theoretical results for two different $I=2$ S-wave
phase-shift representations, the one coming from the ${\cal O}(p^4)$ ChPT result (dashed line) and the one given
by the Schenk parameterization~\cite{Schenk:1991xe} (continuous line). As it can be seen the latter provides a much better agreement
to the data in the low-energy region, however the pass through $\pi/2$ depends little on the parameterization used for $\delta_0^2(s)$.
\par
In summary, we have seen that the zeros of the amplitude of elastic $\pi \pi$ scattering carry important dynamical information on the
structure of the interaction, in particular the role of spin-1 resonances. This information, encoded in the low-energy couplings of
${\cal O}(p^4)$ ChPT, emerges through the zero contour definition and the condition in Eq.~(\ref{eq:master}).
Alternatively the unitarization in Eq.~(\ref{eq:pedrodia}) is able to show the mark of the $\rho(770)$ in the $\delta_1^1(s)$ phase-shift,
as we demonstrated in Figure~\ref{fig:2}. This is not a trivial exercise because the results of ChPT are only valid
for $E \ll M_{\rho}$, however the smoothness of the zero contour and its stability under
unitarization procedures \cite{Arneodo:1973yv,Pennington:1973hi}, collaborate to disentangle the information of the LECs,
at least in the channel we are considering.

\section{The Electroweak Chiral Lagrangian}
\label{sec:3}
In the absence of a Higgs, a strong interacting sector responsible for providing masses to the electroweak gauge bosons is described by Goldstone
bosons $\pi^a, a=1,2,3$, associated to the $SU(2)_L \otimes U(1)_Y \longrightarrow U(1)_{\mbox{em}}$ spontaneous symmetry breaking,
which become the longitudinal components of the electroweak gauge bosons.
The corresponding EChET Lagrangian is then described by the non-linear sigma model based on the coset
$SU(2)_L \otimes SU(2)_R / SU(2)_{L+R}$ where $SU(2)_L \otimes U(1)_Y$ is gauged. The $SU(2)_{L+R} \equiv SU(2)_C$ is
the custodial symmetry that is usually enforced in order
to keep the relation $M_W = M_Z \cos \theta_W$ and the smallness of the $T$ oblique parameter.
\par
A convenient parameterization of the Goldstone fields is given by:
\begin{equation} \label{eq:ux}
 U(x) = \exp \left( \frac{i}{v} \, \pi^a \tau^a \right) \, ,
\end{equation}
with $\tau^a$ the Pauli matrices. This transforms as $L U R^{\dagger}$, with $L \in SU(2)_L$ and $R \in U(1)_Y$, under the gauge group.
Up to dimension four operators, the most general $SU(2)_L \otimes U(1)_Y$ gauge invariant and CP-invariant
Lagrangian which implements the global symmetry breaking
$SU(2)_L \otimes SU(2)_R$ into $SU(2)_{L+R}$ in the limit when $g^\prime$ vanishes\footnote{Recall that the $U(1)_Y$ interactions explicitly
break the global $SU(2)_L \otimes SU(2)_R$ and the custodial $SU(2)_{L+R}$ symmetries.},
is given by the terms
\cite{Longhitano:1980iz,Longhitano:1980tm,Appelquist:1993ka}:
\begin{equation} \label{eq:echet}
 {\cal L}_{\mbox{\footnotesize EChET}} = \frac{v^2}{4} \langle \left( D_{\mu} U \right)^{\dagger} D^{\mu} U \rangle +
\sum_{i=0,\dots 5} \, a_i \, {\cal O}_i \, ,
\end{equation}
with the operators:
\begin{eqnarray} \label{eq:echeto}
{\cal O}_0 &=&  g^{\prime 2} \, \frac{v^2}{4} \langle T \, V_{\mu} \rangle^2 \, , \nonumber \\
{\cal O}_1 &=&  \frac{i g g^\prime}{2} \, B^{\mu\nu} \langle T \, W_{\mu \nu} \rangle \, , \nonumber \\
{\cal O}_2 &=&  \frac{i g^\prime}{2} \, B^{\mu\nu} \langle T \, [V^\mu, V^\nu] \rangle \, , \nonumber \\
{\cal O}_3 &=&  i \,    g \, \langle W_{\mu \nu} \left[ V^{\mu}, V^{\nu} \right] \rangle \, , \nonumber \\
{\cal O}_4 &=&   \langle V_{\mu} V_{\nu} \rangle^2 \, , \nonumber \\
{\cal O}_5 &=&    \langle V_{\mu} V^{\mu} \rangle^2
%{\cal O}_{11} &=&    \langle ( D_{\mu} V^{\mu} )^2 \rangle
\, ,
\end{eqnarray}
where $V_{\mu} = \left( D_{\mu} U \right) U^{\dagger}$, $T=U\,\tau^3\,U^\dagger$. The covariant derivative takes the form:
\begin{equation} \label{eq:dcova}
 D_{\mu} U = \partial_{\mu} U + \frac{i}{2} \, g \, \tau^k \, W_{\mu}^k U - \frac{i}{2}\,  g'\,  \tau^3 \, U B_{\mu} \, ,
\end{equation}
with $W_{\mu \nu} = \tau^k \, W_{\mu \nu}^k / 2$, and being $W_{\mu \nu}^k$ and $B_{\mu \nu}$ the gauge field strength tensors. In Eqs.~(\ref{eq:echet},\ref{eq:echeto}) $\langle ... \rangle$ denotes
the trace in the $SU(2)$ space. Other 8 possible operators, that violate custodial symmetry when $g^\prime \to 0$
(see {\it e.g.}~\cite{Dobado:1997jx}), will not be considered here.
${\cal L}_{\mbox{\footnotesize EChET}}$ involves a
perturbative derivative expansion as ChPT does, driven this time by the scale
$\Lambda_{\mbox{\footnotesize EW}} = 4 \pi v \simeq 3 \, \mbox{TeV}$, {\it i.e.} an expansion in
powers of $(p^2,M_V^2)/\Lambda_{\mbox{\footnotesize EW}}$.

The first term in the Lagrangian~(\ref{eq:echet}) has dimension two, and
provides the SM mass terms for the gauge bosons. The rest of terms
are dimension 4 operators, and only ${\cal O}_{3-5}$ are relevant for the purposes
of our work. ${\cal O}_3$ represents an anomalous triple gauge-boson coupling
while ${\cal O}_4$ and ${\cal O}_5$ give  anomalous quartic gauge-boson interactions.
The corresponding low-energy couplings $a_3$, $a_4$ and $a_5$
encode the information of the heavier spectrum that has been integrated out in order to
get ${\cal L}_{\mbox{\footnotesize EChET}}$. They are the analogous to the chiral LECs $\ell_i^r(\mu)$
discussed in Section \ref{sec:2}.

\subsection{Longitudinally polarized gauge boson scattering}
\label{subsec:31}
In this work we are interested in the scattering of vector bosons with longitudinal polarization
because is the one linked, through the Higgs mechanism, with the Goldstone bosons of the electroweak symmetry breaking sector.
The exact relation is provided by the {\em equivalence theorem}~\cite{Cornwall:1974km,Lee:1977yc}:
\begin{equation} \label{eq:et1}
 A \left( V^a_L V^b_L \rightarrow V^c_L V^d_L \right) =
A \left( \pi^a \pi^b \rightarrow \pi^c \pi^d \right) + {\cal O}\left( \frac{M_V}{E} \right) \, ,
\end{equation}
which states that, at center of mass energies $E\gg M_V$, the amplitude for the elastic
scattering of longitudinally polarized vector bosons ($V_L^a$) equals the amplitude where the gauge fields
have been replaced by their corresponding Goldstone bosons ($\pi^a$).
Now, we can use
the electroweak effective Lagrangian~(\ref{eq:echet}) to calculate the amplitude of Goldstone boson
scattering of Eq.~(\ref{eq:et1}). Since the effective Lagrangian formalism is a low-energy expansion, some care is needed
to apply the equivalence theorem, which is valid in the high-energy limit. The restricted version of the theorem which applies to the
gauge boson scattering amplitude calculated at ${\cal O}(p^4)$ reads~\cite{Dobado:1993dg}:
\begin{eqnarray} \label{eq:et2}
 A \left( V^a_L V^b_L \rightarrow V^c_L V^d_L \right)
&=&
A^{(4)} \left( \pi^a \pi^b \rightarrow \pi^c \pi^d \right)
+ {\cal O}\left( \frac{M_V}{E} \right)
\nonumber \\  &&
+ \, {\cal O}(g,g^\prime) + {\cal O} \left( \frac{E^5}{\Lambda_{\mbox{\footnotesize EW}}^5} \right)\, ,
\end{eqnarray}
where $A^{(4)}$ is the amplitude of Goldstone boson scattering at lowest order in the electroweak
couplings ($g$ and $g^\prime)$ as obtained from the effective Lagrangian
${\cal L}_{\mbox{\footnotesize EChET}}$.
Therefore only the operators ${\cal O}_4$ and  ${\cal O}_5$ in Eq.~(\ref{eq:echeto}) contribute to that amplitude, which is linear
in the $a_4$ and $a_5$ couplings. Let us remark that at ${\cal O}(g^0, g'^0)$ the masses of the gauge bosons
vanish and the equivalence theorem, as given by Eq.~(\ref{eq:et2}), indicates that the Goldstone boson
scattering amplitude has to be calculated in the zero mass limit. Mass corrections appear in the neglected terms.
\par
Notice that the restricted version given by Eq.~(\ref{eq:et2}) is valid in the energy range given by
$M_V \ll E \ll \Lambda_{\mbox{\footnotesize EW}}$.
Since the EChET framework is analogous to ChPT, and the latter works reasonably well
up to $500$~MeV $\simeq 2\pi F_\pi$ , we can assume that the effective formalism for the electroweak theory
is limited at about $2\pi v\simeq 1.5$~TeV. Given the success of the zero-contour method for the case of the $\rho(770)$ resonance,
whose mass is larger than the limit of validity of the theory, the range above may be extended up to $E \lsim 2$~TeV, at least for what concerns the determination of the resonance mass through the zero contours.

\section{Analysis of the zeros of the $W_L Z_L \rightarrow W_L Z_L$ amplitude}
\label{sec:4}
\begin{figure*}[t!]
  \begin{center}
  \includegraphics[width=0.32\textwidth]{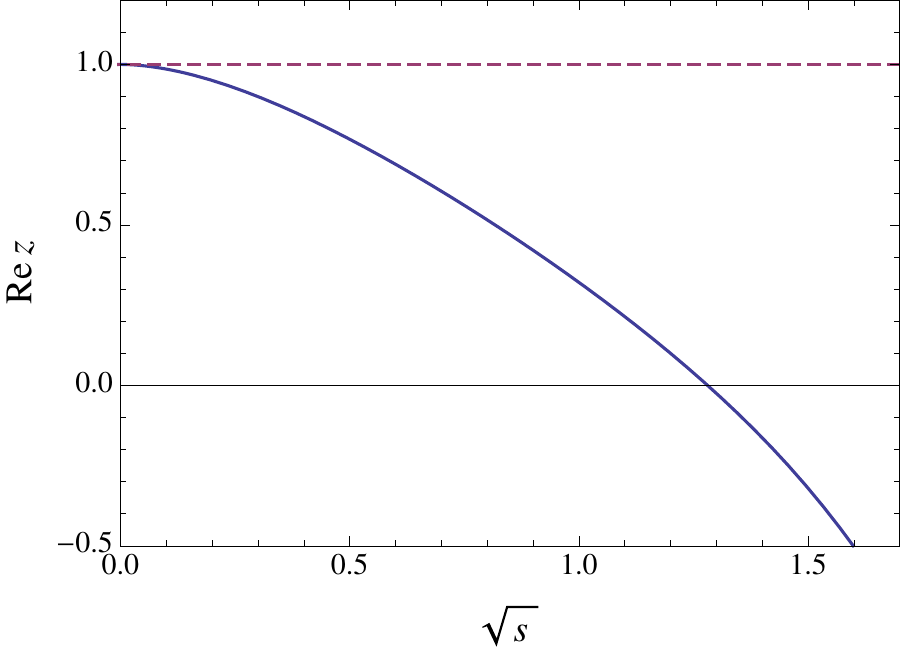}
  \includegraphics[width=0.32\textwidth]{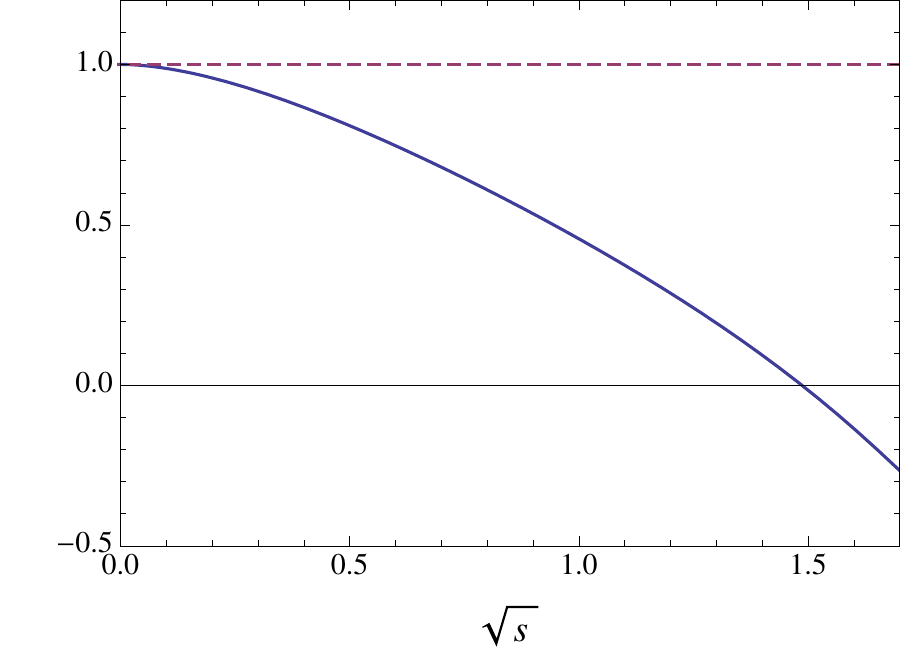}
  \includegraphics[width=0.32\textwidth]{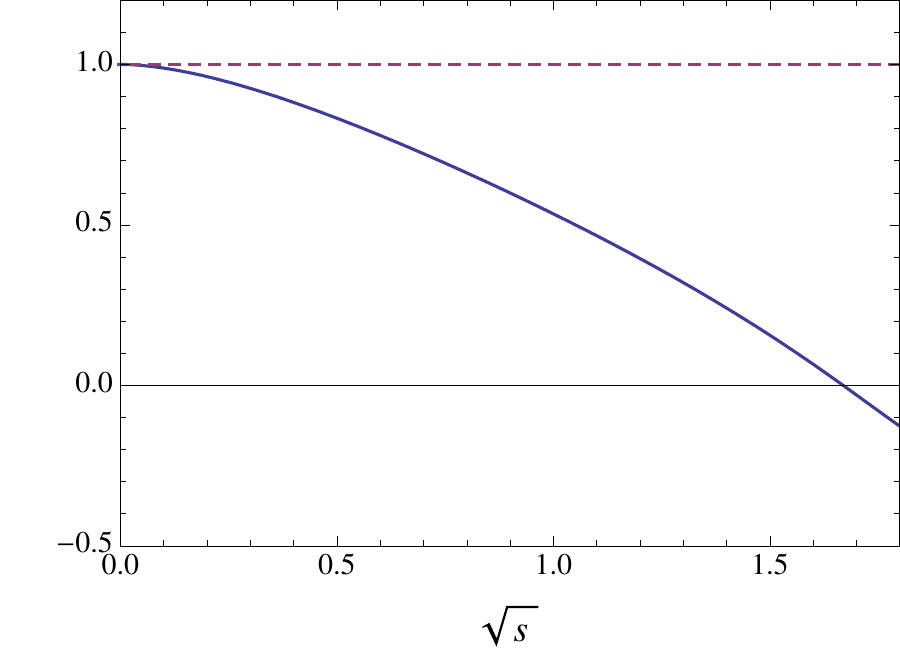}\\ $ \, $ \\
  \includegraphics[width=0.32\textwidth]{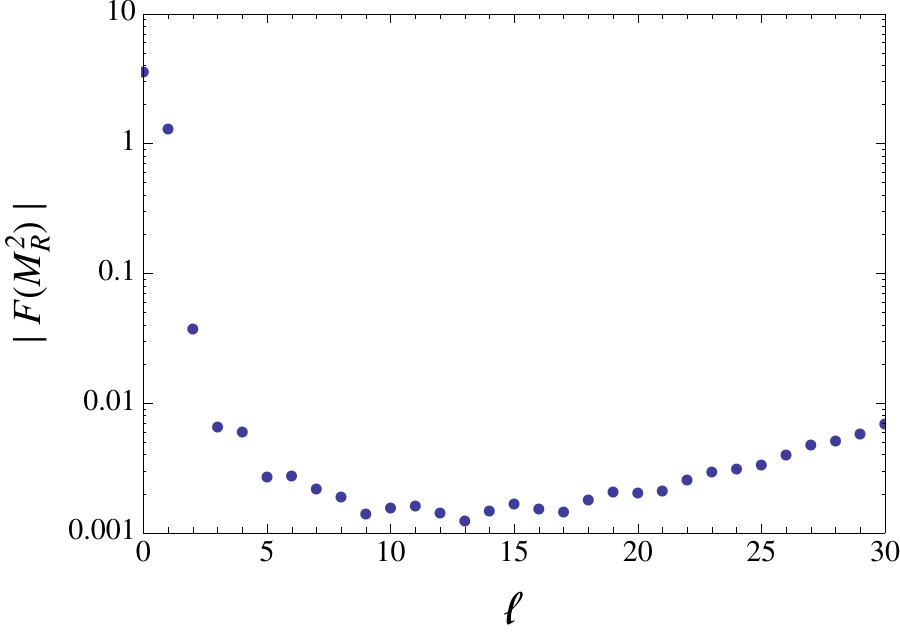}
  \includegraphics[width=0.32\textwidth]{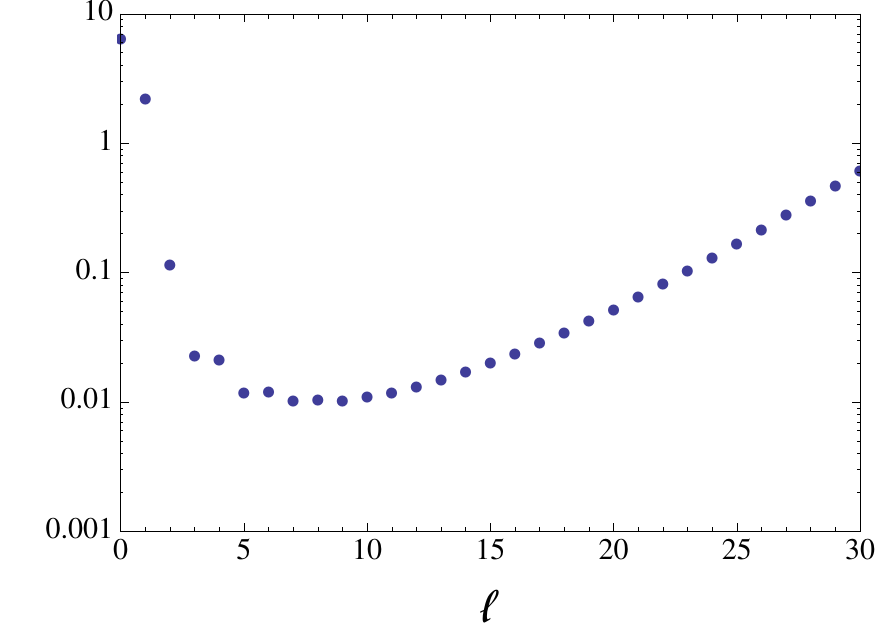}
  \includegraphics[width=0.32\textwidth]{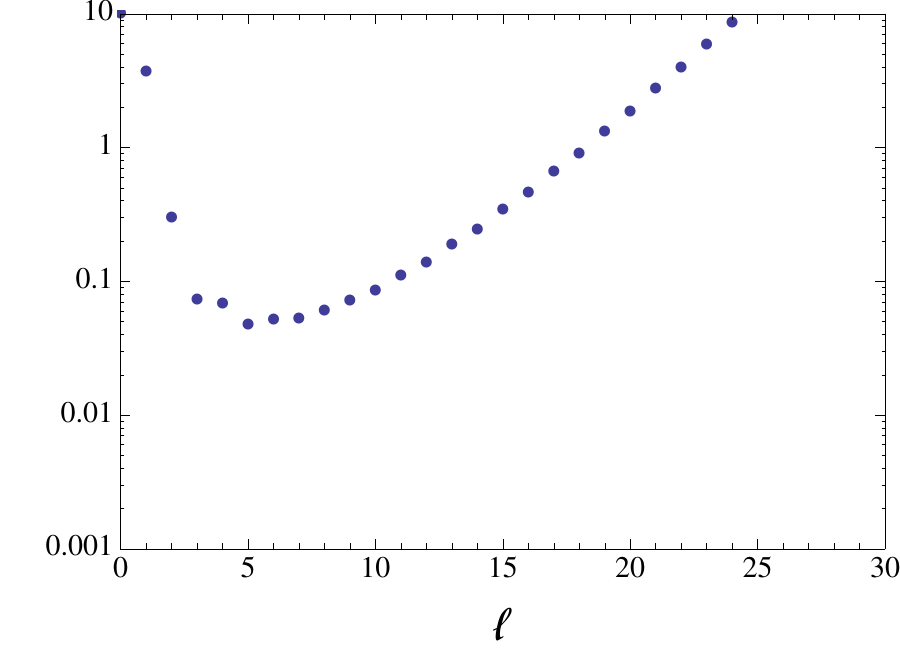}
  \caption{ \label{fig:3}
Upper panels: Zero contours from the ${\cal O}(p^4)$ amplitude for, from left to right,
$(\bar{a}_4,\bar{a}_5)=(10,10)$,
$(\bar{a}_4,\bar{a}_5)=(8.5,10)$ and $(\bar{a}_4,\bar{a}_5)=(7.7,10)$. The dashed line are the amplitude zeros obtained from the
lowest order chiral amplitude.  Lower panels: Corresponding ${\cal O}(p^4)$ amplitudes $|F(M_R^2)| \equiv |A^{(4)}(M_R^2,z_0(M_R^2))|$ (in logarithmic scale) for  the upper panel cases,
using the partial-wave expansion up to  order $\ell$.
 }
  \end{center}
\end{figure*}
The equivalence theorem pointed out in the last section can be used to relate, at leading order, the amplitude of $W_L Z_L \rightarrow W_L Z_L$ with
the one of the corresponding Goldstone bosons, which is analogous to the
$\pi^- \pi^0 \rightarrow \pi^- \pi^0$ amplitude described in Section~\ref{sec:2}. Notice that the physical system provided by the Higgsless Lagrangian
in Eq.~(\ref{eq:echet}) is,
but for the change of scale ($F_{\pi} \rightarrow v$), the same than the one of the ChPT. Therefore one would expect a similar dynamics
if the P-wave contribution is saturated by a vector resonance. Consequently
we could apply the same procedure and study the occurrence of $I=1$ vector resonances~\footnote{Isospin, in the context of the Lagrangian (\ref{eq:echet}), indicates the quantum number associated
to the unbroken $SU(2)_{L+R}$ custodial symmetry.} in the scattering $W_L Z_L \rightarrow W_L Z_L$
through the analysis
of the zero contours of the EChET amplitude.
As explained in Section~\ref{sec:2}, zero contours cross the resonance location
close to where the Legendre polynomial vanishes, which for vector resonances amounts to the condition (\ref{eq:master}).
\par
By virtue of the equivalence theorem, the amplitude for $W_L Z_L \rightarrow W_L Z_L$ is equal, up to ${\cal O}(p^4)$ terms,
to $A^{(4)} \left(\pi^- \pi^0 \rightarrow \pi^- \pi^0 \right)=A(t,s,u)$ in Eq.~(\ref{eq:op4pp}), with the trivial replacements:
$$
 F \to v\quad \; \; ,\quad \; \; (\bar{\ell}_1,\bar{\ell}_2) \to (\bar{a}_5,\bar{a}_4)\,.
$$
In addition, the limit $M \rightarrow 0$ has to be performed, according to the restricted form of the theorem.
This amounts to writing $M=0$ in the polynomial
contributions in $A(t,s,u)$. Care has to be taken in the one-loop functions, where keeping the leading order in that limit
leaves a mass dependence in the logarithms.
The scale-independent $\bar{a}_i$ couplings are related to
their renormalized counterparts in the $\overline{\mbox{MS}}$ scheme as:
\begin{eqnarray} \label{eq:abar}
a_4^r(\mu) &=& \frac{1}{4} \frac{1}{48 \pi^2} \left( \overline{a}_4 -1 + \ln \frac{M_W^2}{\mu^2} \right) \, , \nonumber \\
a_5^r(\mu) &=& \frac{1}{4} \frac{1}{96 \pi^2} \left( \overline{a}_5 -1 + \ln \frac{M_W^2}{\mu^2} \right) .
\end{eqnarray}
This definition for $a_4^r(\mu)$ and $a_5^r(\mu)$ differs from that of Eq.~(\ref{eq:lbar}) that relates the $\ell_i^r(\mu)$ to the $\bar{\ell}_i$;  it matches
though the definition used in recent literature~\cite{Eboli:2006wa,Fabbrichesi:2007ad} for these couplings,
so it is adopted here to make contact with those results. The natural order of magnitude of the couplings is
$\bar{a}_{4,5}\sim {\cal O}(1)$, so we can expect that $a_i^r\sim {\cal O}(10^{-3})$.
\par
Recalling the procedure that we used in Section~\ref{sec:2}, we will look for the zero contours of the amplitude,
$A^{(4)}(s,z_0)=0$, and identify the vector resonances with solutions
of $\mbox{Re}\,z_0(M_R^2)=0$.
However we still need
to impose another constraint on this result to ensure that the assumptions leading to condition (\ref{eq:master}) are fulfilled.
As it was seen in Section~\ref{sec:2}, our procedure depends crucially, after neglecting higher partial waves, on having a dominant
resonance-saturated P-wave and a small, non-vanishing, S-wave contribution.
We can translate this requirement to a numerical bound in the following way. Writing the equation $A^{(4)}(s,z_0)=0$
in terms of its partial-wave series, one gets:
\begin{equation} \label{eq:imb0}
 f_0^2(s) +  3\,f_1^1(s) \, z_0(s) \simeq 0 \, ,
\end{equation}
if higher-order partial waves are neglected. At the resonance location, $s=M_R^2$, the latter equation relates the
size of the imaginary part of the zero  with the ratio between the S- and the P-wave contributions:
\begin{equation} \label{eq:imb}
\left| z_0(M_R^2) \right|  = \left| \mbox{Im} \,z_0(M_R^2) \right| = \left| \frac{f_0^2(M_R^2)}{3 \, f_1^1(M_R^2)} \right| < \lambda \, .
\end{equation}
The bound $\lambda$ then defines the range of applicability of our method: zeros of the amplitude with imaginary part smaller than $\lambda$ can be
 considered positive results in the search for vector resonances. The zeros which pass this condition bear a similarity to the near-by zeros
 introduced in the pioneering works on zero contours, characterized by small imaginary parts.
A reference value for $\lambda$ can be inferred from the $\rho(770)$ case studied in Section~\ref{sec:2},
where one gets $|\mbox{Im} \, z_0(M_{\rho}^2) | \simeq 0.36$
(for $\mu=0.77$~GeV; see Table~\ref{tab:1}). For values of $\lambda$ larger than $1/2$ we cannot consider the S-wave to be significantly smaller than
the P-wave, and we therefore choose $\lambda = 1/2$ as a limiting value for the identification of resonances in the zero contours.
The dependence of our results on this cut is discussed later.
\par
The $\lambda$-cut in Eq.~(\ref{eq:imb}) is also related with the convergence of the partial-wave expansion of the amplitude.
For the zero mass case the partial-wave expansion is convergent only in the physical region, i.e. for
$z \equiv \cos \theta \in [-1,1]$
(see Appendix~\ref{app:A}). The partial-wave series continued to complex values of $z$ is at best
asymptotically convergent. Hence smaller imaginary parts of the zeros imply a better behaviour of the series.
In the lower panels of Figure~\ref{fig:3} we show the convergence of the
partial-wave series at $s=M_R^2$ for three representative examples, corresponding (from left to right)
to $(\bar{a}_4,\bar{a}_5)=(10,10), (8.5,10), (7.7,10)$, with non-zero imaginary parts
given by
$|\mbox{Im}\, z_0(M_R^2)| \simeq 0.26, 0.39$ and $0.56$, respectively.
As can be seen there is a direct correlation between the size of the imaginary part
of $z_0(M_R^2)$ and the convergence of the expansion. Note also that the minimal value of the amplitude
at the resonance location gets closer to zero for smaller $|\mbox{Im}\, z_0(M_R^2)|$.
The corresponding zero contours, obtained from the amplitude~(\ref{eq:op4pp}) (now keeping only the leading order in the $M \rightarrow 0$
limit), are smooth lines in the $(\mbox{Re}\,z,s)$ plane, as shown in the upper panels of Figure~\ref{fig:3}. The dashed line are the amplitude zeros obtained from the
lowest order chiral amplitude, which are simply given by $t=0$ and thus lie on a straight line in the Mandelstam plane. In terms of
$z$ the ${\cal O}(p^2)$ zeros are given by $z_0(s)=1$
and have no imaginary part because the ${\cal O}(p^2)$ amplitude is real. The ${\cal O}(p^4)$ corrections induce an imaginary part in the zero
trajectories,
and are essential to bend down the zero contour towards Re~$z=0$. The crossing, from left to right in
Figure~\ref{fig:3}, takes place at $M_R\simeq 1.28, 1.49$ and $1.67$~TeV.
\begin{figure*}[t!]
\begin{center}
\includegraphics[width=13cm]{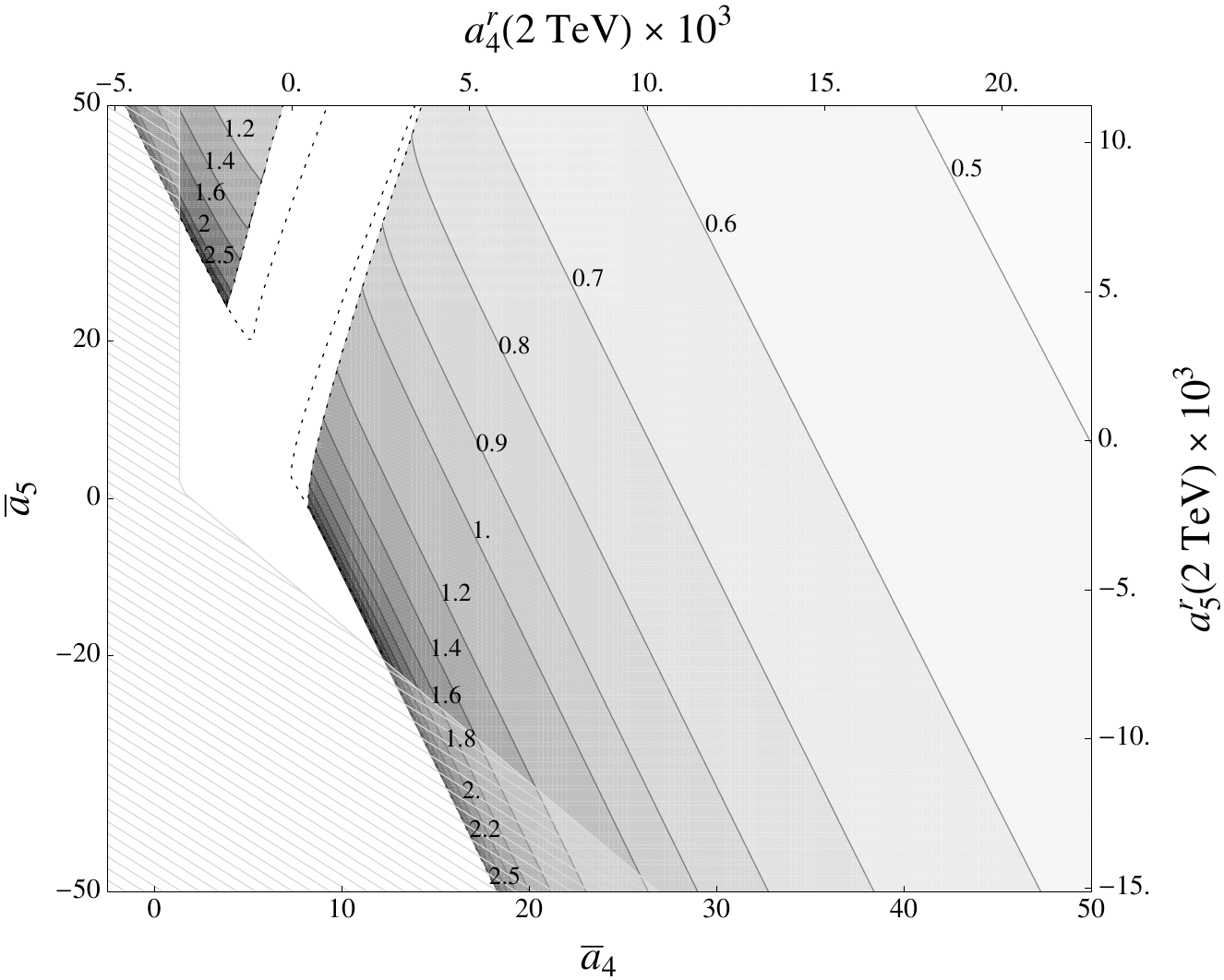}
\caption{\label{fig:4} Resonance masses as a function of the low-energy couplings $\bar{a}_4$ and $\bar{a}_5$.
 The scales in terms of
the renormalized couplings $a_4^r(\mu)$ and $a_5^r(\mu)$ at $\mu=2$~TeV are also drawn.
The shaded areas show where resonances defined by the conditions~(\ref{eq:master}) and~(\ref{eq:imb})
with $\lambda=1/3$ are found in the $(\bar{a}_4,\bar{a}_5)$-plane. The contour lines drawn correspond to pairs of $(\bar{a}_4,\bar{a}_5)$ which yield the same resonance mass. The hatched region in the left and lower parts of the plot, given by Eq.~(\ref{eq:Imz}), corresponds to values of
$\bar{a}_4$ and $\bar{a}_5$ forbidden by positivity conditions on the $\pi\pi$ scattering amplitudes.
The outermost dashed lines mark the boundary of the resonance region corresponding to $\lambda=1/2$.}
\end{center}
\end{figure*}
\par
Figure~\ref{fig:4} is the central result of our work. The shaded areas show where resonances, defined by the conditions (\ref{eq:master})
and (\ref{eq:imb}) with $\lambda = 1/3$, are found in the $(\bar{a}_4,\bar{a}_5)$-plane.
The contour lines drawn correspond to pairs of $(\bar{a}_4,\bar{a}_5)$ which yield the same resonance mass. Though the validity
of the approach cannot be trusted beyond $E \simeq 2$~TeV, we have displayed in the plot resonances found with masses up to $2.5$ TeV.
In order to show how dependent are the solutions from Eq.~(\ref{eq:master}) to the cut~(\ref{eq:imb}) on the imaginary part of the
zeros, we have also drawn in Figure~\ref{fig:4} (outer dashed line) the boundary of the region yielding resonances
when $\lambda= 1/2$.
The hatched region in the left and lower parts of the plot corresponds to values of
$\bar{a}_4$ and $\bar{a}_5$ forbidden by positivity conditions on the $\pi\pi$ scattering amplitudes.
These bounds were obtained in Ref.~\cite{Pennington:1994kc}, and slightly improved in Ref.~\cite{Manohar:2008tc}.
Translated to $\bar{a}_4$ and $\bar{a}_5$ they read:
\begin{eqnarray} \label{eq:Imz}
\bar{a}_5 + 2 \,\bar{a}_4 \ge \frac{157}{40} \quad,\quad \bar{a}_4 \ge \frac{27}{20} \,.
\end{eqnarray}
\par
Let us comment the most relevant features of the results of Figure~\ref{fig:4}:
\begin{itemize}
\item[i/] No vector resonances are found for $\bar{a}_4\lesssim 8$ and $\bar{a}_5\lesssim 25$.
This would exclude to a large extent
Higgsless models with vector resonances which saturate the low-energy couplings to the
expected natural order of magnitude ($\bar{a}_{4,5}\sim 1$).
\item[ii/] Masses above $1.8$~TeV are confined to a thin slice in the lower-left and upper-left parts of the shaded regions
and are mostly excluded by the positivity constraints. Conversely, light resonances ($\lesssim 0.8$~TeV) require values of either $\bar{a}_4$ or $\bar{a}_5$ larger than 20. The validity of the EChET Lagrangian for such large values of the LECs is nevertheless questionable and could indicate that additional degrees of freedom linked to the Goldstone boson dynamics are missing in the effective description.
\item[iii/] The dependence on the $\lambda$-cut is visible in the upper part ($\bar{a}_5\gsim 0$)
of the boundary of the allowed region for resonances, where the $\lambda=1/2$ contour departs
from the $\lambda= 1/3$ one, which defines the shaded region. Making the cut smaller would further constrain the
region where resonances are found. We nevertheless think that the value
$\lambda=1/3$ is a realistic one, since it provides a reasonable suppression of S-waves and it is also found in the
QCD case for the $\rho(770)$ resonance.
\item[iv/] For resonance masses $M_R\lesssim 0.8\, \mbox{TeV}$, where either $\bar{a}_4$ or $\bar{a}_5$ are large, the formula:
\begin{eqnarray} \label{eq:approxMR}
M_R = v \left(  \frac{ 192\, \pi^2 }{ -11+10\,\bar{a}_4+2\,\bar{a}_5} \right)^{1/2} \,,
\end{eqnarray}
provides an approximation to the resonance masses in Figure~\ref{fig:4} with an accuracy better than 10\%. The
formula (\ref{eq:approxMR}) corresponds to the zero contours obtained by setting the loop functions $\bar{J}(x)$ in $A(t,s,u)$ to zero,
and further neglecting the mass $M$.
\item[v/] A final consistency check for our method is provided by the Eq.~(\ref{eq:pedrodia}) for the $I=1$ P-wave phase-shift $\delta_1^1$.
We evaluated the energy at which $\delta_1^1 = \pi\slash 2$ for each value of the parameters $\bar{a}_4$, $\bar{a}_5$ in the shaded region
of Figure~\ref{fig:4}
and we found that the result is basically the same we obtained from the condition in Eq.~(\ref{eq:master}).
Indeed, if we allow for at most a 10\% deviation between both mass determinations, only the points
in a tiny slice lying exactly on the boundary of the lower-half part of the big shaded region
(in fact mostly excluded already by the positivity constraints) fail to pass the test. This result confirms that the S-wave background
is indeed small, and therefore that the condition Re~$z_0(M_R^2)\simeq 0$ is a characteristic signature for vector resonances.
As it was commented in relation with Figure~\ref{fig:2}, this result does not depend, essentially, on our knowledge of the
 $\delta_0^2(s)$ phase shift. It is enough that
this is small with respect to the $\delta_1^1(s)$ phase shift at $E \simeq M_R$.
\end{itemize}

The LHC sensitivity to explore the values of the coefficients $a_4$ and $a_5$ has been investigated in
Ref.~\cite{Eboli:2006wa}. The reported limits imply that in the combined region $\bar{a}_4 \lsim 35$ and
$ -38 < \bar{a}_5 < 45$ no deviation from the SM prediction could be observed at the LHC.
The prospects of measuring these parameters with improved accuracy in a high-luminosity
$e^+e^-$ collider operating at 1~TeV are slightly better~\cite{Boos:1999kj}.
On the other hand, the present bounds on new neutral vector resonances obtained
recently~\cite{Eboli:2011ye}  using ATLAS and CMS data, which exclude masses up to 1-2.3 TeV
depending on their couplings and widths, could be used in combination with our results in Figure~\ref{fig:4}
to constrain the allowed regions of $\bar{a}_4$ and $\bar{a}_5$ which
can accommodate a vector-saturated model.
\par
A systematic study of resonance masses in the parametric space spanned by $a_4^r(\mu)$ and $a_5^r(\mu)$ using the Inverse Amplitude Method has 
also been performed~\cite{Dobado:1999xb,Pelaez:1996wk}. Their results, compared with our Figure~\ref{fig:4}, look rather different. 
From the pole of the Pad\'e-improved P-wave they find for the vector resonance masses the result:
\begin{equation} \label{eq:mtheirs}
 M_R = v \left( \frac{144 \, \pi^2}{3 \, \bar{a}_4- 3 \, \bar{a}_5 + 1} \right)^{1/2} \, ,
\end{equation}
that can be compared with our Eq.~(\ref{eq:approxMR}), though the latter is only valid for large values of $\bar{a}_4, \bar{a}_5$, 
i.e.  for $M_R \lesssim 0.8\, \mbox{TeV}$. The slope of the lines of equal mass do not agree. Moreover, formula~(\ref{eq:mtheirs})
forbids vector resonances in the region defined by $\bar{a}_5 > \bar{a}_4+ 1/3$, where their existence is consistent with our
conditions. Also we note that the results of~\cite{Dobado:1999xb,Pelaez:1996wk} predict resonances in the region of $\bar{a}_{4,5}\sim 1$,
that contradict our findings. 
A detailed comparative analysis between both methods in order to trace the origin of the discrepancies shall be carried out elsewhere. 

\section{Conclusion}
\label{sec:5}
Under the assumption that no light SM Higgs will be found at the LHC, we have investigated a
method to identify vector resonances originated from a strong electroweak symmetry-breaking sector in the 1~TeV energy region.
These resonances have become a possible alternative to the SM Higgs in preventing the seeming loss of perturbative
 partial-wave unitarity in the elastic scattering of the longitudinal components of W and Z gauge bosons.
More important is that those resonances would provide a clear signal that a strong interacting
dynamics is responsible for the spontaneous breaking of the electroweak symmetry.
\par
We have focused, in particular, on the resonances that could contribute to the $W_L Z_L \rightarrow W_L Z_L $ scattering. This channel
has the appropriate characteristics to implement our approach, that searches for vector resonances dominating the amplitude.
Assuming resonance saturation of the LECs of the effective chiral Lagrangian 
describing the interaction among Goldstone bosons, we could extract relevant information on the lightest vector resonances from the zeros of
the elastic scattering amplitude.
We first applied our method to the well known case of the $\rho(770)$ resonance and
the $\mathcal{O}(p^4)$ chiral $\pi^-\pi^0\to \pi^-\pi^0$ amplitude
and considered the impact of introducing the next order in the chiral expansion.
\par
Turning to the electroweak case we exploited the fact that, at leading order in the expansion provided by the equivalence theorem,
the dynamics of the longitudinally polarized
gauge boson scattering is described by the electroweak chiral
Lagrangian~(\ref{eq:echet}), which is identical to
the one that generates the $\rho(770)$ in elastic $\pi \pi$ scattering upon the
obvious change of scale $F_{\pi} \rightarrow v$.
Within this approach we have explored the parameter space of the two low-energy couplings $\bar{a}_4$ and $\bar{a}_5$
needed to describe the $W_L Z_L \rightarrow W_L Z_L $ scattering amplitude, in order to identify the region where a vector
resonance can dominate the amplitude, and provide an estimate of its mass. The outcome has
been shown in
Figure~\ref{fig:4} as a contour plot in the ($\bar{a}_4$,$\bar{a}_5$)-plane. Our main conclusion
is that no vector resonances are found for $\bar{a}_4\lesssim 8$ and $\bar{a}_5\lesssim 25$, indicating that Higgsless models with
vector resonances which saturate the low-energy couplings to the expected natural order of magnitude ($\bar{a}_{4,5}\sim 1$) would be excluded
 to a large extent. If we consider the neighborhood outside that natural order
of magnitude, we see that the first resonances,
appearing for $\bar{a}_{4} \gsim 8$, have masses above $1 \, \mbox{TeV}$.
Lighter vector resonance masses appear for rather
unnatural values of the parameters.
\par
The improvement of our method by the inclusion of the ${\cal O}(p^6)$ contributions will entail a few technical
difficulties,
since it carries many new couplings
(as it was seen in the
QCD case in Section~\ref{sec:2}) that make an analogous treatment to the one proposed here rather cumbersome.
Nevertheless, if LHC confirms that there is no light Higgs, the study of the zeros of the gauge boson scattering amplitudes with the use of
effective field theories driven by the spontaneous symmetry breaking pattern
could provide
a model-independent tool to explore the role of the resonances emerging from the new strong dynamics
and shall therefore be pursued, for instance, through the search for higher spin states.

\section*{Acknowledgements}
We wish to thank M.R.~Pennington and A.~Pich for comments and suggestions on the writing of this
article, A.~Pich for conversations on this topic and G.~Ecker for his insight on the chiral
${\cal O}(p^6)$ LEC estimates.
This project is partially supported by MEC (Spain) under grants
FPA2007-60323 and FPA2011-23778, by the Spanish Consolider-Ingenio 2010 Programme CPAN
(CSD2007-00042) and by Generalitat Valenciana under grant
PROMETEO/2008/069.

\appendix
\renewcommand{\theequation}{\Alph{section}.\arabic{equation}}
\renewcommand{\thetable}{\Alph{section}.\arabic{table}}
\setcounter{equation}{0}
\setcounter{table}{0}

\section{Convergence of the partial-wave expansion in elastic $\pi \pi$ scattering}
\label{app:A}

The partial-wave expansion of the elastic $\pi\pi$ scattering amplitude
for a fixed $s$ is defined in the real interval $z=[-1,1]$, but it can be analytically continued to a larger region in
the complex $z$-plane, according to a well-known theorem by K.~Neumann~(see {\it e.g.} Ref.~\cite{davis}).
The theorem states that the expansion of a function $f(z)$ in a series of Legendre polynomials is absolutely convergent in the interior
 of the largest ellipse with foci at $z=\pm 1$ in which $f(z)$ is analytic, and divergent in the exterior of the ellipse.
In our case $f(z)$ is the ${\cal O}(p^4)$ chiral amplitude $A(t(s,z),s,u(s,z))$, Eq.~(\ref{eq:op4pp}),
which for a fixed $s$ has branch discontinuities
 at the $t$-channel and $u$-channel thresholds, {\it i.e.} at $t=4M^2$ and $u=4M^2$. In the complex $z$-plane this
translates to branch cuts  extending from $z_+=1+8M^2/(s-4M^2)$ to $+\infty$ and from  $(-z_+)$ to $-\infty$.
The region of convergence of the partial-wave expansion for $A(t(s,z),s,u(s,z))$ is thus
an ellipse with foci at $z=\pm 1$ and semi-major axis $z_+$.
The semi-minor axis of the
ellipse of convergence is equal to
\begin{eqnarray} \label{eq:semiminor}
z_- = \sqrt{z_+^2 -1 } = \frac{ 4\,M\, \sqrt{s}   }{ s-4 M^2 } \,,
\end{eqnarray}
and limits the size of the imaginary part of $z$ for which the partial-wave series converges.
Incidentally, for $M=0$ the ellipses contracts to the interval $z=[-1,1]$ and, hence, the partial-wave expansion is
not convergent for $\mbox{Im}\, z \neq 0$.

%\bibliographystyle{utphys_alb}
%\bibliography{refzero}

\begin{thebibliography}{60}

\bibitem{Veltman:1976rt}
M.~Veltman,
{\em Acta Phys. Polon.} {\bf B8} (1977)  475.
%%CITATION = APPOA,B8,475;%%.

\bibitem{Lee:1977yc}
B.~W. Lee, C.~Quigg, and H.~Thacker,
\href{http://dx.doi.org/10.1103/PhysRevLett.38.883}{{\em Phys. Rev. Lett.} {\bf
  38} (1977)  883--885}.
%%CITATION = PRLTA,38,883;%%.

\bibitem{Veltman:1991cm}
H.~G. Veltman and M.~Veltman,
{\em Acta Phys. Polon.} {\bf B22} (1991)  669--696.
%%CITATION = APPOA,B22,669;%%.

\bibitem{Weinberg:1978kz}
S.~Weinberg,
{\em Physica} {\bf A96} (1979)  327.
%%CITATION = PHYSA,A96,327;%%.

\bibitem{Gasser:1983yg}
J.~Gasser and H.~Leutwyler,
\href{http://dx.doi.org/10.1016/0003-4916(84)90242-2}{{\em Annals Phys.} {\bf
  158} (1984)  142}.
%%CITATION = APNYA,158,142;%%.

\bibitem{Ecker:1988te}
G.~Ecker, J.~Gasser, A.~Pich, and E.~de~Rafael,
\href{http://dx.doi.org/10.1016/0550-3213(89)90346-5}{{\em Nucl. Phys.} {\bf
  B321} (1989)  311}.
%%CITATION = NUPHA,B321,311;%%.

\bibitem{Pennington:1973dz}
M.~Pennington,
{\em AIP Conf. Proc.} {\bf 13} (1973)  89--116.
%%CITATION = APCPC,13,89;%%.

\bibitem{Pennington:1994kc}
M.~Pennington and J.~Portol\'es,
\href{http://dx.doi.org/10.1016/0370-2693(94)01551-M}{{\em Phys. Lett.} {\bf
  B344} (1995)  399--406}.
%%CITATION = HEP-PH/9409426;%%.

\bibitem{Cornwall:1974km}
J.~M. Cornwall, D.~N. Levin, and G.~Tiktopoulos,
\href{http://dx.doi.org/10.1103/PhysRevD.10.1145, 10.1103/PhysRevD.11.972}{{\em
  Phys. Rev.} {\bf D10} (1974)  1145}.
%%CITATION = PHRVA,D10,1145;%%.

\bibitem{Longhitano:1980iz}
A.~C. Longhitano,
\href{http://dx.doi.org/10.1103/PhysRevD.22.1166}{{\em Phys. Rev.} {\bf D22}
  (1980)  1166}.
%%CITATION = PHRVA,D22,1166;%%.

\bibitem{Longhitano:1980tm}
A.~C. Longhitano,
\href{http://dx.doi.org/10.1016/0550-3213(81)90109-7}{{\em Nucl. Phys.} {\bf
  B188} (1981)  118}.
%%CITATION = NUPHA,B188,118;%%.

\bibitem{Appelquist:1993ka}
T.~Appelquist and G.-H. Wu,
\href{http://dx.doi.org/10.1103/PhysRevD.48.3235}{{\em Phys. Rev.} {\bf D48}
  (1993)  3235--3241}.
%%CITATION = HEP-PH/9304240;%%.

\bibitem{MartinSpearman:1970}
A.~Martin and T.~Spearman, {\em {Elementary Particle Theory}} (North-Holland
  Pub. Co., Amsterdam, 1970)  .

\bibitem{MartinMorgan:1976}
B.~Martin, D.~Morgan, and G.~Shaw, {\em {Pion-Pion Interactions in Particle
  Physics}} (Academic Press, London, 1976)  .

\bibitem{Dobado:1997jx}
A.~Dobado, A.~G\'omez-Nicola, A.~L. Maroto, and J.~Pel\'aez,
{\em {Effective lagrangians for the standard model}} (Springer, Heidelberg,
  1997)  .
%%CITATION = INSPIRE-455294;%%.

\bibitem{Pennington:1971fd}
M.~Pennington and P.~Pond,
\href{http://dx.doi.org/10.1007/BF02823325}{{\em Nuovo Cim.} {\bf A3} (1971)
  548--560}.
%%CITATION = NUCIA,A3,548;%%.

\bibitem{Pennington:1973xv}
M.~Pennington and S.~Protopopescu,
\href{http://dx.doi.org/10.1103/PhysRevD.7.1429}{{\em Phys. Rev.} {\bf D7}
  (1973)  1429--1441}.
%%CITATION = PHRVA,D7,1429;%%.

\bibitem{Pennington:1973hi}
M.~Pennington and C.~Schmid,
\href{http://dx.doi.org/10.1103/PhysRevD.7.2213}{{\em Phys. Rev.} {\bf D7}
  (1973)  2213--2219}.
%%CITATION = PHRVA,D7,2213;%%.

\bibitem{Odorico:1972eb}
R.~Odorico,
\href{http://dx.doi.org/10.1016/0370-2693(72)90169-4}{{\em Phys. Lett.} {\bf
  B38} (1972)  411--414}.
%%CITATION = PHLTA,B38,411;%%.

\bibitem{Barrelet:1971pw}
E.~Barrelet,
\href{http://dx.doi.org/10.1007/BF02732655}{{\em Nuovo Cim.} {\bf A8} (1972)
  331--371}.
%%CITATION = NUCIA,A8,331;%%.

\bibitem{Pennington:1972zp}
M.~Pennington and S.~Protopopescu,
\href{http://dx.doi.org/10.1016/0370-2693(72)90296-1}{{\em Phys. Lett.} {\bf
  B40} (1972)  105--108}.
%%CITATION = PHLTA,B40,105;%%.

\bibitem{Gasser:1984gg}
J.~Gasser and H.~Leutwyler,
\href{http://dx.doi.org/10.1016/0550-3213(85)90492-4}{{\em Nucl. Phys.} {\bf
  B250} (1985)  465}.
%%CITATION = NUPHA,B250,465;%%.

\bibitem{Pich:2010sm}
A.~Pich, I.~Rosell, and J.~J. Sanz-Cillero,
\href{http://dx.doi.org/10.1007/JHEP02(2011)109}{{\em JHEP} {\bf 1102} (2011)
  109}.
%%CITATION = ARXIV:1011.5771;%%.

\bibitem{Dumm:2009va}
D.~G\'omez~Dumm, P.~Roig, A.~Pich, and J.~Portol\'es,
\href{http://dx.doi.org/10.1016/j.physletb.2010.01.059}{{\em Phys. Lett.} {\bf
  B685} (2010)  158--164}.
%%CITATION = ARXIV:0911.4436;%%.

\bibitem{Weinberg:1966kf}
S.~Weinberg,
\href{http://dx.doi.org/10.1103/PhysRevLett.17.616}{{\em Phys. Rev. Lett.} {\bf
  17} (1966)  616--621}.
%%CITATION = PRLTA,17,616;%%.

\bibitem{Bijnens:1997vq}
J.~Bijnens, G.~Colangelo, G.~Ecker, J.~Gasser, and M.~Sainio,
\href{http://dx.doi.org/10.1016/S0550-3213(97)00621-4,
  10.1016/S0550-3213(97)00621-4}{{\em Nucl. Phys.} {\bf B508} (1997)  263--310}.
%%CITATION = HEP-PH/9707291;%%.

\bibitem{Bijnens:1999hw}
J.~Bijnens, G.~Colangelo, and G.~Ecker,
\href{http://dx.doi.org/10.1006/aphy.1999.5982}{{\em Annals Phys.} {\bf 280}
  (2000)  100--139}.
%%CITATION = HEP-PH/9907333;%%.

\bibitem{Gasser:2009hr}
J.~Gasser, C.~Haefeli, M.~A. Ivanov, and M.~Schmid,
\href{http://dx.doi.org/10.1016/j.physletb.2009.03.056}{{\em Phys. Lett.} {\bf
  B675} (2009)  49--53}.
%%CITATION = ARXIV:0903.0801;%%.

\bibitem{Cirigliano:2006hb}
V.~Cirigliano, G.~Ecker, M.~Eidem\"uller, R.~Kaiser, A.~Pich, and J.~Portol\'es,
\href{http://dx.doi.org/10.1016/j.nuclphysb.2006.07.010}{{\em Nucl. Phys.} {\bf
  B753} (2006)  139--177}.
%%CITATION = HEP-PH/0603205;%%.

\bibitem{Protopopescu:1973sh}
S.~Protopopescu, M.~Alston-Garnjost, A.~Barbaro-Galtieri, S.~M. Flatte,
  J.~Friedman, {\em et al.},
\href{http://dx.doi.org/10.1103/PhysRevD.7.1279}{{\em Phys. Rev.} {\bf D7}
  (1973)  1279}.
%%CITATION = PHRVA,D7,1279;%%.

\bibitem{Ochs:1973}
W.~Ochs, {\em Thesis submitted to the University of Munich} (1973)  .

\bibitem{Estabrooks:1974vu}
P.~Estabrooks and A.~D. Martin,
\href{http://dx.doi.org/10.1016/0550-3213(74)90488-X}{{\em Nucl. Phys.} {\bf
  B79} (1974)  301}.
%%CITATION = NUPHA,B79,301;%%.

\bibitem{Gasser:1990ku}
J.~Gasser and U.~G. Meissner,
\href{http://dx.doi.org/10.1016/0370-2693(91)91235-N}{{\em Phys. Lett.} {\bf
  B258} (1991)  219--224}.
%%CITATION = PHLTA,B258,219;%%.

\bibitem{Schenk:1991xe}
A.~Schenk,
\href{http://dx.doi.org/10.1016/0550-3213(91)90236-Q}{{\em Nucl. Phys.} {\bf
  B363} (1991)  97--116}.
%%CITATION = NUPHA,B363,97;%%.

\bibitem{Arneodo:1973yv}
A.~Arneodo, F.~Guerin, and J.~Donohue,
\href{http://dx.doi.org/10.1007/BF02777939}{{\em Nuovo Cim.} {\bf A17} (1973)
  329--342}.
%%CITATION = NUCIA,A17,329;%%.

\bibitem{Dobado:1993dg}
A.~Dobado and J.~Pel\'aez,
\href{http://dx.doi.org/10.1016/0550-3213(94)90174-0,
  10.1016/0550-3213(94)90174-0}{{\em Nucl. Phys.} {\bf B425} (1994)  110--136}.
%%CITATION = HEP-PH/9401202;%%.

\bibitem{Eboli:2006wa}
O.~\'Eboli, M.~Gonz\'alez-Garc\'{\i}a, and J.~Mizukoshi,
\href{http://dx.doi.org/10.1103/PhysRevD.74.073005}{{\em Phys. Rev.} {\bf D74}
  (2006)  073005}.
%%CITATION = HEP-PH/0606118;%%.

\bibitem{Fabbrichesi:2007ad}
M.~Fabbrichesi and L.~Vecchi,
\href{http://dx.doi.org/10.1103/PhysRevD.76.056002}{{\em Phys. Rev.} {\bf D76}
  (2007)  056002}.
%%CITATION = HEP-PH/0703236;%%.

\bibitem{Manohar:2008tc}
A.~V. Manohar and V.~Mateu,
\href{http://dx.doi.org/10.1103/PhysRevD.77.094019}{{\em Phys. Rev.} {\bf D77}
  (2008)  094019}.
%%CITATION = ARXIV:0801.3222;%%.

\bibitem{Boos:1999kj}
E.~Boos, H.~He, W.~Kilian, A.~Pukhov, C.~Yuan, {\em et al.},
\href{http://dx.doi.org/10.1103/PhysRevD.61.077901}{{\em Phys. Rev.} {\bf D61}
  (2000)  077901}.
%%CITATION = HEP-PH/9908409;%%.

\bibitem{Eboli:2011ye}
O.~\'Eboli, J.~Gonz\'alez-Fraile, and M.~Gonz\'alez-Garc\'{\i}a,
  \href{http://dx.doi.org/10.1103/PhysRevD.85.055019}{{\em Phys. Rev.} {\bf D85}
  (2012)  055019}.
10 pages, 6 figures.
%%CITATION = ARXIV:1112.0316;%%.

\bibitem{Dobado:1999xb}
A.~Dobado, M.~Herrero, J.~Pel\'aez, and E.~Ruiz~Morales,
\href{http://dx.doi.org/10.1103/PhysRevD.62.055011}{{\em Phys. Rev.} {\bf D62}
  (2000)  055011}.
%%CITATION = HEP-PH/9912224;%%.

\bibitem{Pelaez:1996wk}
J.~Pel\'aez,
\href{http://dx.doi.org/10.1103/PhysRevD.55.4193}{{\em Phys. Rev.} {\bf D55}
  (1997)  4193--4202}.
%%CITATION = HEP-PH/9609427;%%.

\bibitem{davis}
P.~J. {Davis}, {\em Interpolation and Approximation}.
\newblock Dover Publications, New York, first~ed., 1975.

\end{thebibliography}
%\providecommand{\href}[2]{#2}\begingroup\raggedright

%\endgroup

\end{document}